\documentclass[aps,pre,twocolumn,showpacs,groupedaddress,amsmath,amssymb]{revtex4-1}

\usepackage{graphicx}
\usepackage{dcolumn}
\usepackage{bm}
\usepackage{color}
\usepackage[utf8]{inputenc}
\usepackage[bottom]{footmisc}
\usepackage[T1]{fontenc}
\usepackage{float,enumitem}
\usepackage{mathptmx}
\usepackage{subcaption}
\begin{document}

\title[Simulations of multicomponent relativistic thermalization]{Simulations of multicomponent relativistic thermalization}

\author{Atul Kedia}
 \email{akedia@nd.edu, atulkedia93@gmail.com}
 \affiliation{Center for Astrophysics, Department of Physics, University of Notre Dame, Notre Dame, Indiana 46556, USA}
\author{Nishanth Sasankan}%
 \affiliation{Center for Astrophysics, Department of Physics, University of Notre Dame, Notre Dame, Indiana 46556, USA}

\author{Grant J. Mathews}
 \email{gmathews@nd.edu}
 \affiliation{Center for Astrophysics, Department of Physics, University of Notre Dame, Notre Dame, Indiana 46556, USA}
 
\author{Motohiko Kusakabe}
 \affiliation{IRCBBC$,$ School of Physics$,$ Beihang University$,$ Beijing 100083 China}

\date{\today}

\begin{abstract}

Multicomponent relativistic fluids have been studied for decades. However, simulating the dynamics of the particles and fluids in such a mixture has been a challenge due to the fact that such simulations are computationally expensive in three spatial dimensions. Here, we report on the development and application of a multi dimensional relativistic Monte Carlo code to explore of the thermalization process in a relativistic multicomponent environment in a computationally inexpensive way. As an illustration we simulate the fully relativistic three dimensional Brownian-motion-like solution to the thermalization of a high mass particle (proton) in a bath of relativistic low-mass particles (electrons). We follow the thermalization and ultimate equilibrium distribution of the Brownian-like particle as can happen in the cosmic plasma during big bang nucleosynthesis. We also simulate the thermalization of energetic particles injected into the plasma as can occur, for example, by the decay of massive unstable particles during the big bang.
\end{abstract}

\maketitle

\section{Introduction}
Understanding the solution to the Boltzmann equation for fluids is of fundamental importance for its practical implications in chemistry, biophysics, astrophysics, and cosmology. Over the past decades considerable progress has been made toward understanding these solutions for multicomponent mixtures and in relativistic environments\cite{Cubero,DunkelHanggi,1+1,1+3,Kremer, Cercignani}. However, until recently there has been little progress in solving the relativistic multicomponent Boltzmann equation. Existing analytic solutions are based on the nondegenerate limit of the relativistic Boltzmann equation \cite{Sasankan20}, or the Fokker-Planck equation approximation to the relativistic Boltzmann equation \cite{1+3}. While a one dimensional numerical simulation exists and supports the theory \cite{Cubero}, there currently exists no numerical simulation supporting the theory in fully three spatial dimensions (3D). Here we present a Monte Carlo simulation built to replicate the fully relativistic multicomponent Boltzmann equation via a stochastic random walk process. Such a tool should have widespread applications in the dynamics of mixtures of relativistic and mildly-relativistic fluids.

\subsection{Background}
Currently, multicomponent relativistic simulations have been performed only in 1 spatial dimension (1D) where one can vary the number density of each species \cite{Cubero}. However, for three dimensions only approximate analytical and numerical solutions for the relativistic Boltzmann equation currently exist. These are based upon various interpretations of the stochastic process for solving the corresponding Fokker-Planck equation \cite{1+3, Sasankan20, Acosta}. 

Cubero et al. \cite{Cubero} have discussed the difficulty in simulating multi-species thermalization in two and three dimensions. The difficulty being in modelling the complete electromagnetic fields due to all particles in space. This, can be simplified in 1 dimension by treating particles as only undergoing point-like elastic collisions. However, if one applies this simplification in 2 or 3 dimensions the collision probability becomes vanishingly small even when including finite cross sections. This increases the computational time for particles to equilibrate. In this paper, however, we present a new Monte Carlo scheme which mimics the thermalization process from the perspective of one particle undergoing many successive collisions and thus can be studied in minimal computational time. The Monte-Carlo thermalization code can be accessed at \cite{github}.

\subsection{Cosmological application}
As an illustration, we consider here an application to the thermalization of baryons during the epoch of big bang nucleosynthesis (BBN). The relativistic thermalization simulation described here has recently been applied in Ref.~\cite{Sasankan20} to describe the equilibrium kinetic energy distribution of baryons in the BBN environment.

BBN occurs during an epoch of the early universe that lasts from about 1 s to a few minutes and is responsible for the synthesis of light nuclei such as $^2$H, $^3$He, $^4$He, and $^7$Li from pre-existing neutrons and protons remaining after the weak reactions fall out of equilibrium.

For the most part BBN involves two-body nuclear reactions. In this case, each pair of nuclei is directly related to their distribution function in relative velocity. That is, the reaction rate $R_{(1 + 2 \to 3 + ..)}$ between two species is given by
\begin{equation}
    R_{(1 + 2 \to 3 + ..)} = n_1 n_2 <\sigma v> = n_1 n_2 \int \sigma v f(v) d v~~,
\end{equation}
where $n_1$ and $n_2$ are the number densities of colliding nuclei, $\sigma$ is the cross section, $v$ is the relative velocity between the two nuclei, and $f(v)$ is the relative velocity distribution. This distribution is determined from the individual velocity distributions. Among other things, the Monte Carlo simulation described here aims to generate the individual velocity distribution from first principles. This is then applied to various test cases.

At the start of BBN nuclei are immersed in a bath of highly relavitistic electrons, positrons and photons. During BBN the universe expands and cools from a temperature of $kT \approx 1$ MeV to $kT \approx 0.01$ MeV. During this time frame the electron-positron asymmetry begins to manifest as the temperature falls below the electron rest mass ($0.511$ MeV). Initially, the electron number density is orders of magnitude higher than the baryon number density (See Table 1 from Sasankan \cite{Sasankan20}). Even though photons have a high number density w.r.t. baryons ($n_b/n_\gamma \sim 10^{-9}$), they have a low cross section for nuclear scattering. Hence, electron scattering dominates. This implies that nuclei obtain thermal equilibrium, by elastically scattering almost exclusively with mildly relativistic electrons in the cosmic plasma.

A motivation for the present work is that there has been considerable recent interest in the possibility of a modification of the baryon distribution function from Maxwell-Boltzmann (MB) statistics. This modification can be in the form of Tsallis statistics \cite{Tsallis, Kusakabe19,Bertulani13, Hou17}, the influence of inhomogeneous primordial magnetic fields on baryons \cite{Luo19}, non-ideal plasma effects at low temperature \cite{Jang18}, the injection of nonthermal particles (e.g. \cite{Jedamzik04,Kawasaki05,Jedamzik06, Kusakabe09, Cyburt10, Voronchev12, Kusakabe14} and Refs.~therein), and small relativistic corrections to the MB distribution that arise due to nuclear kinetic drag \cite{McDermott18}.

Another case in which a relativistic nonthermal distribution function can emerge during the early universe is in the reheating epoch near the end of cosmic inflation \cite{Harigaya, Mukaida}. As the universe enters the reheating phase the scalar field responsible for driving inflation begins to decay into radiation and particles. It has been argued, however, that thermalization may not occur until well after reheating. This is because the low density and high expansion rate cause the time scale of thermalization to be so long that it delays the evolution of the produced nonthermal particles toward an effective temperature \cite{Mukaida}. The resulting temperature of the universe might then be significantly altered from that obtained under the assumption of instantaneous thermalization. However, in \cite{Harigaya} for example it has been argued that thermalization might still occur via small-angle scatterings. Clearly, this is a case where a detailed Monte Carlo solution to the evolution of the particle distribution functions in an expanding space-time could shed light.

Yet another case that we study here explicitly is the effect of injected nonthermal particles due, for example, to energetic hadronic decays by relic massive (possibly supersymmetric) particles formed during an earlier epoch. As hadrons are injected into the primordial plasma one must follow their evolution along with the baryon distribution functions and their time-dependent effects on the thermonuclear reaction rates. Thus, it is worthwhile to develop a fully relativistic method to describe the time-dependent evolution toward thermalization within the BBN environment.

To demonstrate the viability of this Monte Carlo technique, we here apply our method to several test cases. The cosmological environment poses a good test environment as one component (the baryons) is much heavier than the other (relativistic electrons and photons). Also, as the background temperature changes the lighter particles transition from being relativistic to non-relativistic. This provides a test case which includes regimes where heavy particles are submerged in either a relativistic or non-relativistic bath and are thermalized by electron collisions.

\subsection{The Monte Carlo simulation}
In this paper we describe a Monte Carlo simulation that replicates the thermalization of charged nuclei in a background relativistic fluid. As an illustration, we first follow the thermalization of a proton with zero initial momentum in a bath of relativistic electrons. The simulation obeys general physics conservation laws, including fully relativistic elastic scattering dynamics, and endeavors to mimic how nuclei would exchange energy with it's surroundings. In principle, the nuclear distribution obtained during and until the end of thermalization would be the physical distribution contributing to nuclear reaction rates. As a second test case we follow the time evolution toward thermalization of an injected relativistic 10 GeV proton in the primordial plasma. 

In a sense, this simulation provides an exact solution to the multicomponent relativistic Boltzmann equation by a sequence of elastic scattering events in the same way that Nature does. The Boltzmann equation for the one-particle distribution functions ($f_a$) characterizes collisions of constituent $a$ with constituent particles $b$. This can be written,
\begin{equation}
p^\alpha_a \partial_\alpha f_a = \sum_{b = 1}^r \int{(f_a' f_b' - f_a f_b) F_{ba} \sigma_{ab} d\Omega \frac{d^3p_b}{p_{b0}}} ~~,
\label{Boltz}
\end{equation}
where the right-hand side is the one-particle collision term. The quantity $F_{ba} =\sqrt{(p^\alpha_a p_{b \alpha})^2 - m_a m_b } $ is the invariant flux, while for our purposes $\sigma_{ba}$ is the invariant differential elastic scattering cross section into an element of solid angle $ d\Omega$ that characterizes the collision of constituent $a$ with constituents $b$.

In Sec. II we discuss the algorithm we have developed for simulating this process. That is followed in Sec. III by numerical results we obtain for the illustrative case of protons with zero initial momentum in a bath of relativistic electrons as would be the case in BBN. In Section IV we describe the evolution of the distribution function of a relativistic 10 GeV proton injected by decay into the primordial plasma. We discuss conclusions in Sec. V. In Appendix A we outline the Lorentz-transformations of the distribution functions utilized in the Monte Carlo simulations.

\section{Method}

The Monte Carlo technique we have developed simulates the response of a test particle to numerous elastic scattering events with the background particles. For the first case considered here the test particle is a light nucleus as encountered in BBN. However, the particle mass and scattering cross section with the background species can be modified to study any other physical environment of interest. The background particles in this illustration are electrons and positrons for BBN, and similarly their mass and cross section with the test particle can be modified to study any other particle bath of interest. In this paper, the terms ``test particle" and the ``nucleus" are used interchangeably, as are the terms ``background particle" and ``electron". As noted above we can assume that the test particle scatters predominantly with the background species. This corresponds to an environment in which the test particle number density is much lower than that of the background particles as is the case during BBN. However, this restriction can easily be lifted to simulate more general fluids and plasmas.

\subsection{Initial conditions for the algorithm}
For the illustration considered here we adopt the following initial conditions:
\begin{enumerate}[label=(\alph*)]
    \item The temperature of the electron gas is set to values between $kT =1$ MeV to $0.01$ MeV.
    
    The BBN era starts when the temperature of the universe is about 1 MeV and stops when the universe cools down to 0.01 MeV.
    \item The mass and charge of the nucleus is set to that of the proton (i.e. $Z=1$, $m_p =939$ MeV).
    \item The mass of the electron is set to $m_e = 0.511$ MeV.
    \item The initial total relativistic energy of the test particle nucleon is $E = \sqrt{m^2 c^4 + p^2 c^2} = m + (\gamma-1)m$. For the example of a nucleon initially at rest $E = mc^2 = 939$ MeV. For the example descibed in Section IV of a relativistic injected proton we set $E= 10939$ MeV corresponding to $(\gamma-1)m = 10$ GeV of initial relativistic kinetic energy.
    
\end{enumerate}
From the initial state, we then evolve up to $10^7$ scattering events. This is because, for the most part we notice that for light nuclei during BBN a stationary distribution function is usually obtained after about $3\times 10^6$ scattering events. That is, in the simulations the nucleus eventually experiences a sufficient number of collisions that its initial state is ``forgotten" and becomes irrelevant. Nevertheless, the time it takes to reach the equilibration distribution is of interest here as the nucleus maintains a non-equilibrium distribution prior to the equilibration time.

\subsection{The algorithm with details and reasoning}
\begin{figure*}
    \begin{subfigure}[b]{0.4\textwidth}
        \includegraphics[width=0.8\textwidth]{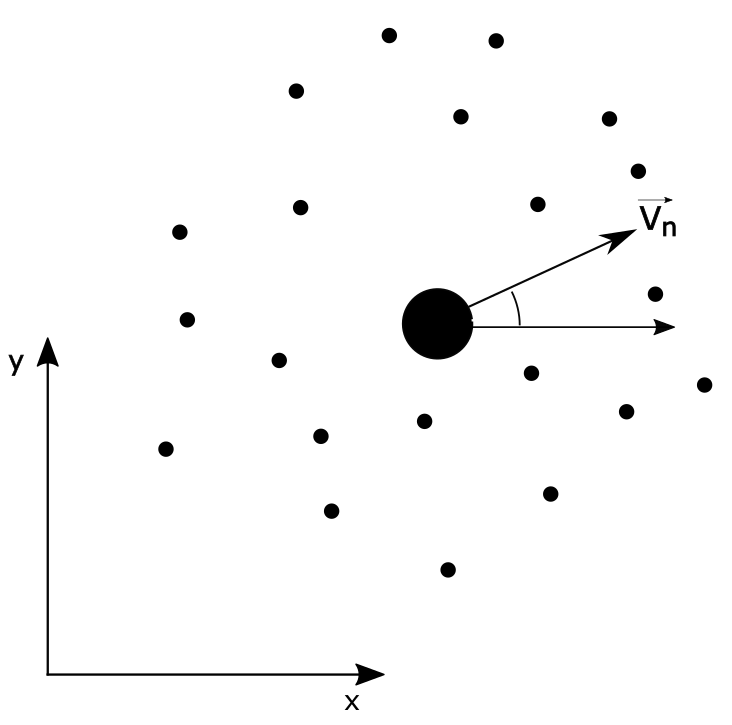}
        \caption{Cosmic rest frame}
    \end{subfigure}
    \begin{subfigure}[b]{0.4\textwidth}
        \includegraphics[width=0.8\textwidth]{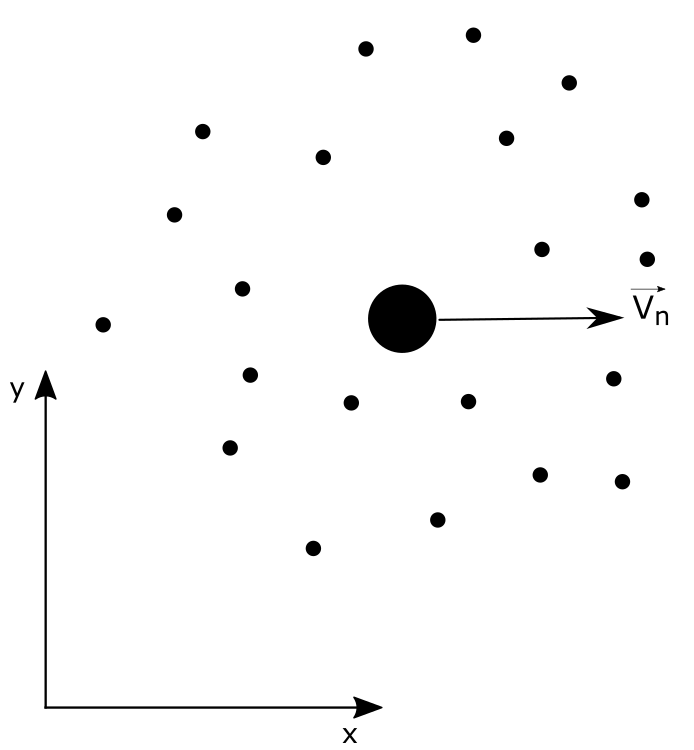}
        \caption{Background rest frame rotated to make $V_n$ in x-direction.}
    \end{subfigure}
    
    \begin{subfigure}[b]{0.4\textwidth}
        \includegraphics[width=0.8\textwidth]{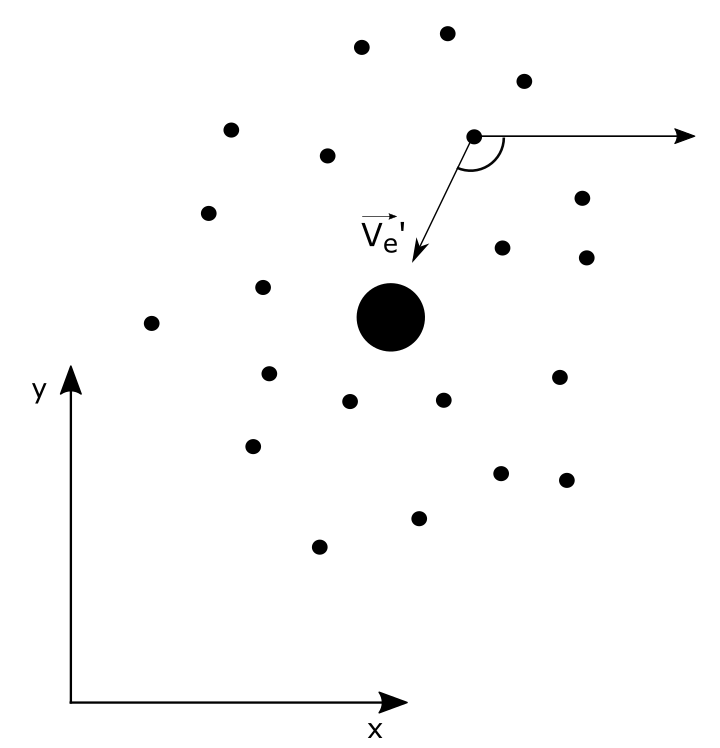}
        \caption{Nucleus frame after Lorentz boosting, with the electron that is about to scatter from the nucleus next.}
    \end{subfigure}
    \begin{subfigure}[b]{0.4\textwidth}
        \includegraphics[width=0.8\textwidth]{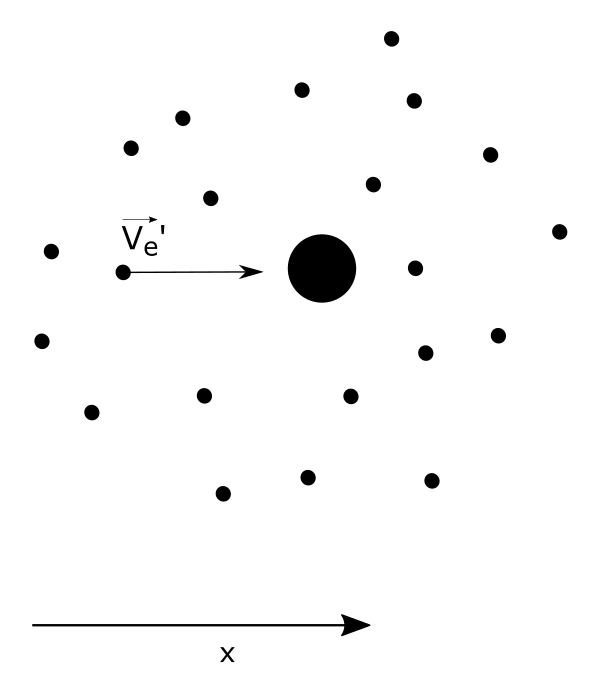}
        \caption{Nuclear frame after rotating to have $V_e$ arrive along the x-direction.}
    \end{subfigure}
    
    \begin{subfigure}[b]{0.2\textwidth}
        \includegraphics[width=0.8\textwidth]{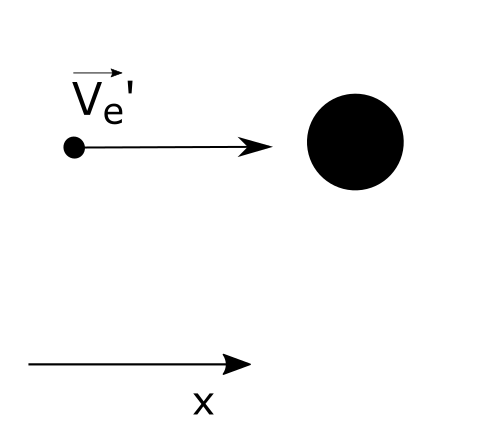}
        \caption{The electron approaching the nucleus in the nuclear rest frame.}
    \end{subfigure}\qquad
    \begin{subfigure}[b]{0.2\textwidth}
        \includegraphics[width=0.8\textwidth]{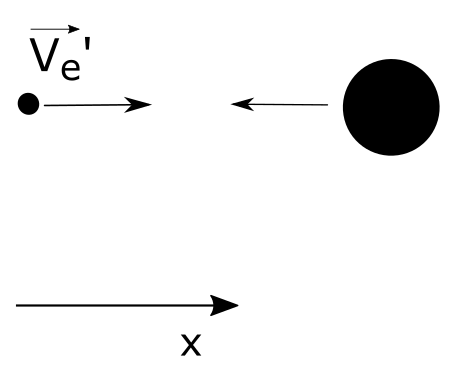}
        \caption{Lorentz boosting to the Center of Momentum(COM) frame.}
    \end{subfigure}\qquad
    \begin{subfigure}[b]{0.2\textwidth}
        \includegraphics[width=0.8\textwidth]{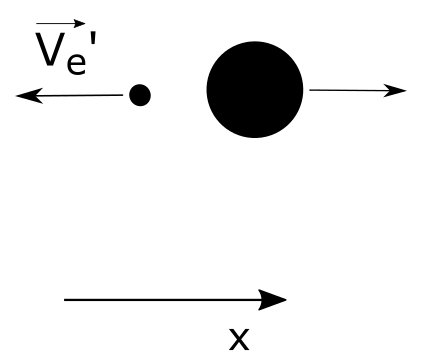}
        \caption{The electron and nucleus after collision in the COM frame.}
    \end{subfigure}
    \caption{Schematic illustration of the progression of rotations and transformations performed until the collision from (a) to (g). Following (g) we Lorentz-transform the nucleus back to the original background fluid frame.}\label{fig:collision}
\end{figure*}

The algorithm to describe the scattering of an electron from the nucleus involves multiple rotations and Lorentz transformations so that the collision parameters are easier to acquire. These steps are schematically illustrated in Figure \ref{fig:collision} a-g. The detailed steps of the algorithm are as follows:
\begin{enumerate}
    \item The simulation starts in the background rest frame.
    
    This is the frame where the collective background momentum is zero.
    \item We rotate the frame to have the velocity of the nucleus be along the $+x$-axis.
    
    We do this to simplify the collision mechanics. This rotation does not affect the background due to the isotropy of the background.
    \item We next make a Lorentz boost to the co-moving frame of the nucleus.
    
    We can then calculate the velocity-dependent flux distribution of the electrons approaching the nucleus. This samples the electron that will interact with the nucleus next.
    \item We determine the electron distribution in the co-moving frame using the derivation described in the Appendix.
    
    This is obtained by applying number conservation between the moving and rest frame, finding volume element conversion, followed by converting all variables into their corresponding Lorentz-transformed value. In 3-D, the Fermi-Dirac (FD) distribution representative of the background fluid in a boosted frame is given by:
    \begin{align}
     f_{FD,3D}'(\mathbf{v'}) &= (\frac{1}{n})\frac{\gamma_o}{\gamma'_V} \gamma'^5 \frac{1}{\Big(1+\exp\Big(\frac{\gamma'\gamma_V(1+Vv_x')mc^2}{kT}\Big)\Big)}
    \end{align}
    Here, ${1}/{n}$ is a normalization constant, $\gamma'_V$ and $\gamma_o$ are the Lorentz factors for the speed of the frames, i.e. they should be $\gamma_o = 1$ (for the cosmic frame, which is at rest with respect to the background cloud) and $\gamma'_V = {1}/{\sqrt{1-V^2}}$ (where $V$ is the speed of the boosted frame). $\mathbf{v'}$ is the background electron velocity and $\gamma' = {1}/{\sqrt{1-v'^2}}$. $f_{FD,3D}'$ is the velocity distribution in 3D in the boosted frame, i.e. the rest frame of the nucleus.
    \item We select an electron randomly based upon this distribution. Specifically we choose the electron's velocity vector from the incoming flux rate
    \begin{align}
        R(\theta) \sim v' f_{FD,3D}'(\mathbf{v'}) ~~.
        \label{vf(v)}
    \end{align}
    This electron will be the one that scatters off the nucleus for this iteration and in the process changes the nuclear four-momentum.
    
    The electron bath surrounds the nucleus in all directions. The angular part of the distribution of electron velocity depicts the fraction of electrons moving in each direction. We select an electron velocity from the distribution using a Monte Carlo technique. The direction of the velocity is the direction in which the electron approaches the nucleus starting from an arbitrary distance away.
    \item We rotate the frame such that the electron approaches the nucleus from the ($-$) $x$-direction and is moving with a positive velocity $v_x$.
    
    Once the electron that collides with a nucleus is chosen, we ignore the rest of the background and this rotation makes it easier to describe the elastic scattering.
    \item We Lorentz transform to the center of momentum (COM) frame of the nucleus-electron system.
    
    Moving to the center of momentum (COM) simplifies the collision. In this frame, the nucleus and electron approach with equal and opposite 3-momenta. When the elastic collision happens, the total 4-momentum is conserved.
    \item We select the scattering angle from the angular distribution.
    
    To determine the scattering angle the differential Mott cross section's angular distribution is used. The differential Mott cross section formula is given by,
    \begin{equation}
            \frac{d\sigma}{d\cos\theta} = \frac{\pi Z^2 \alpha^2}{2 v^2p^2 \sin^4 \frac{\theta}{2}} \left(1 - \frac{v^2}{c^2} \sin^2 \frac{\theta}{2} \right) ~~,
        \label{Mott}
    \end{equation}
    where, $\theta$ is the scattering angle, $\alpha$ is the fine structure constant, $Z$ is the nuclear charge, and $v$ and $p$ are the velocity and momentum of the electron.
    Given a unique incoming electron velocity, the differential scattering cross section is a distribution of probabilities of the various scattering angles based upon the impact parameter and thus obeys the scattering angular distribution.
    
    Note that the Mott cross section and its non-relativistic counterpart the Rutherford cross section are singular at a zero scattering angle, i.e. at large impact parameters. Individually these grazing scatterings, however, do not contribute significantly towards exchanging momentum between the interacting particles. Thus, they do not contribute significantly towards the thermalization in limited computational time. To avoid having to simulate these numerous but insignificant grazing scattering cases, we restrict our simulations to impact parameters less than 3 times the proton radius, i.e. $\sim 2.5$ fm and calculate the corresponding minimum scattering angle using the relation,
    \begin{equation}
            b = \frac{Z_1 Z_2e^2}{4\pi\epsilon_0mv^2}\cot\bigg(\frac{\theta}{2}\bigg) ~~.
        \label{impact_parameter}
    \end{equation}
    Hence, our scattering angles range from this minimum angle to a maximum of 180 degree corresponding to a head-on collision.
    
    We have also made a simplified version of the Monte Carlo simulation that only allows head-on collisions, and hence, does not employ the Mott differential cross section for scattering angle selection. We notice that the resultant equilibration rate and equilibrium distribution are nearly identical to those obtained when including the Mott cross section as described here. Hence, this option can be used as in \cite{Sasankan20} to minimize computation time.
    
    \item We calculate the final electron and nucleus momenta based upon the scattering angle and the initial incoming momentum of the electron.
    
    The momentum calculations are done using four-momentum conservation. However, for the head on only case in 3D and 2D and in the 1D case there are only head-on collisions so that the momenta of the two particles simply gets exchanged.
    
    Once the collision is completed, the electron is no longer considered. The electron moves away from the nucleus and under the assumption of molecular chaos does not interact with the nucleus again. Hence, it is irrelevant and can be ignored.
    
    \item From here we transform the nucleus back to the background rest frame.
    
    The transformations that follow are performed to obtain the velocity and energy of the nucleus in the background rest frame. That is we: 
    \item Lorentz transform the velocity of the nucleus back to the precollision rest frame of the nucleus.
    \item Rotate the velocity of the nucleus to have scattering along the direction the electron was initially approaching.
    \item Lorentz transform the velocity of the nucleus to the background rest frame.
    \item Repeat from the beginning of the algorithm with this (moving) nucleus as the test particle.
\end{enumerate}

For our test cases, repeated scattering between the nucleus and electrons is sufficient to produce the distribution the nucleus attains during BBN. However, for more complicated fluids, involving more than one background species one can easily expand this simulation technique to include scattering events from the different species onto one test particle. This can be done by adding scattering events from the other species and carefully selecting the incoming particle based upon the reaction rate of the test particle with each of the background species.

\section{Resulting distribution for a proton in a relativistic electron bath}
We first tested our scattering algorithm in a simulation to obtain the equilibrium thermalization for a proton initially at rest as the nucleus in an electron bath with various fixed temperatures relevant to the BBN environment. The procedure was as described in Sec. II. We performed separate simulations for different temperatures from the onset of BBN at $kT = 1$ MeV to the conclusion of BBN at $kT = 0.01$ MeV. For the most part, the thermal equilibration occurs at a rate faster than the cosmic expansion timescale \cite{McDermott18}. Hence, for this first illustration it is adequate to approximate the thermalization of nuclei at fixed constant temperatures. We also note, that since we are only concerned with the final equilibrium distribution for this illustration, the choice of a particle initially at rest is arbitrary. We have also considered cases of the proton injected with initially 0.001 MeV or 10 MeV and even 10 GeV (see below) initially and the resultant final thermalized distributions are indistinguishable. We also note that for this illustration we do not consider the electron chemical potential since for most of BBN the chemical potential is negligible. This is because the $e^+-e^-$ symmetry is not completely broken until near the end of BBN at $kT\sim 0.01$ MeV.

\begin{figure}[H]
\includegraphics[height=2.3in,width=3.5in]{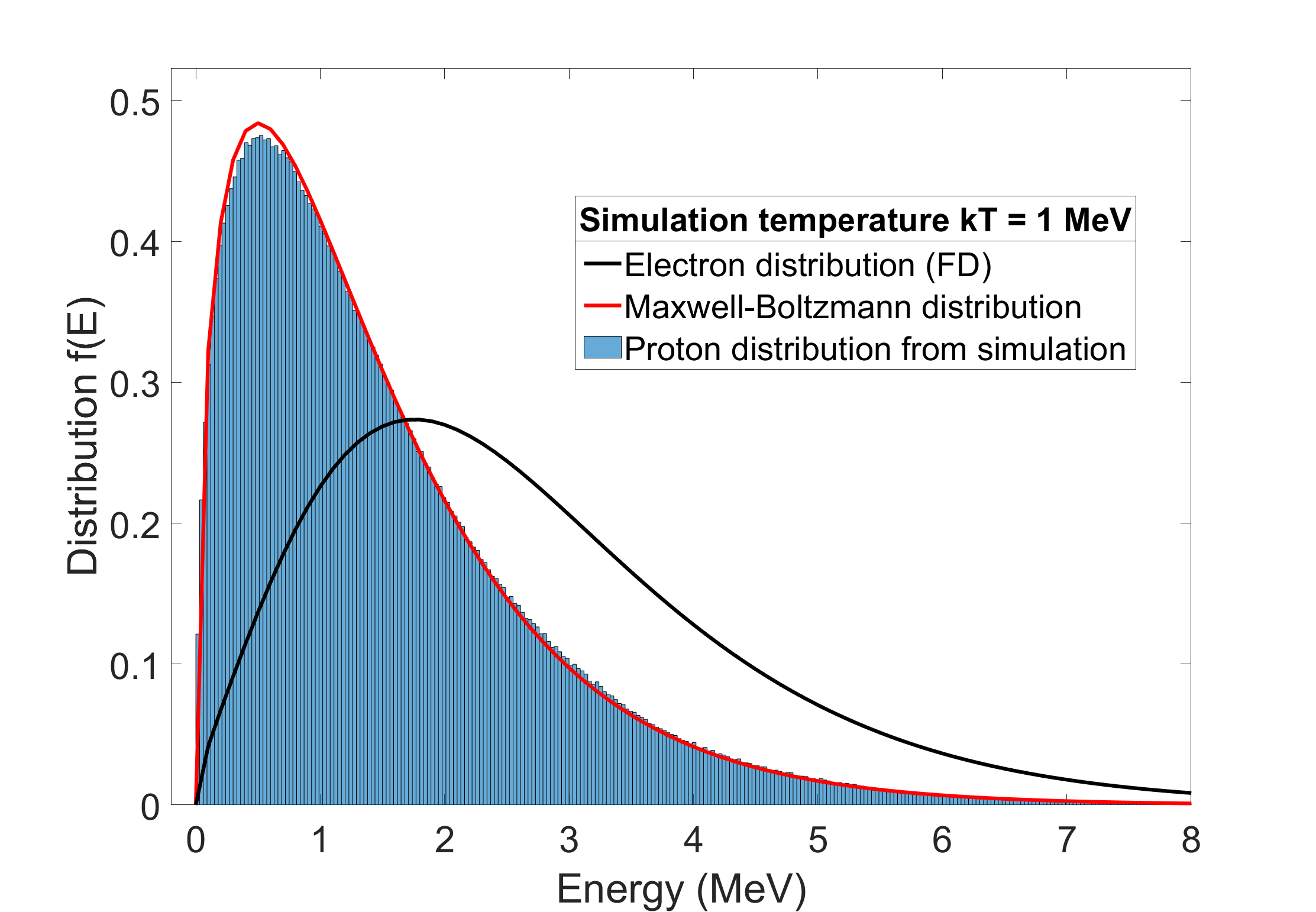}
\includegraphics[height=2.3in,width=3.5in]{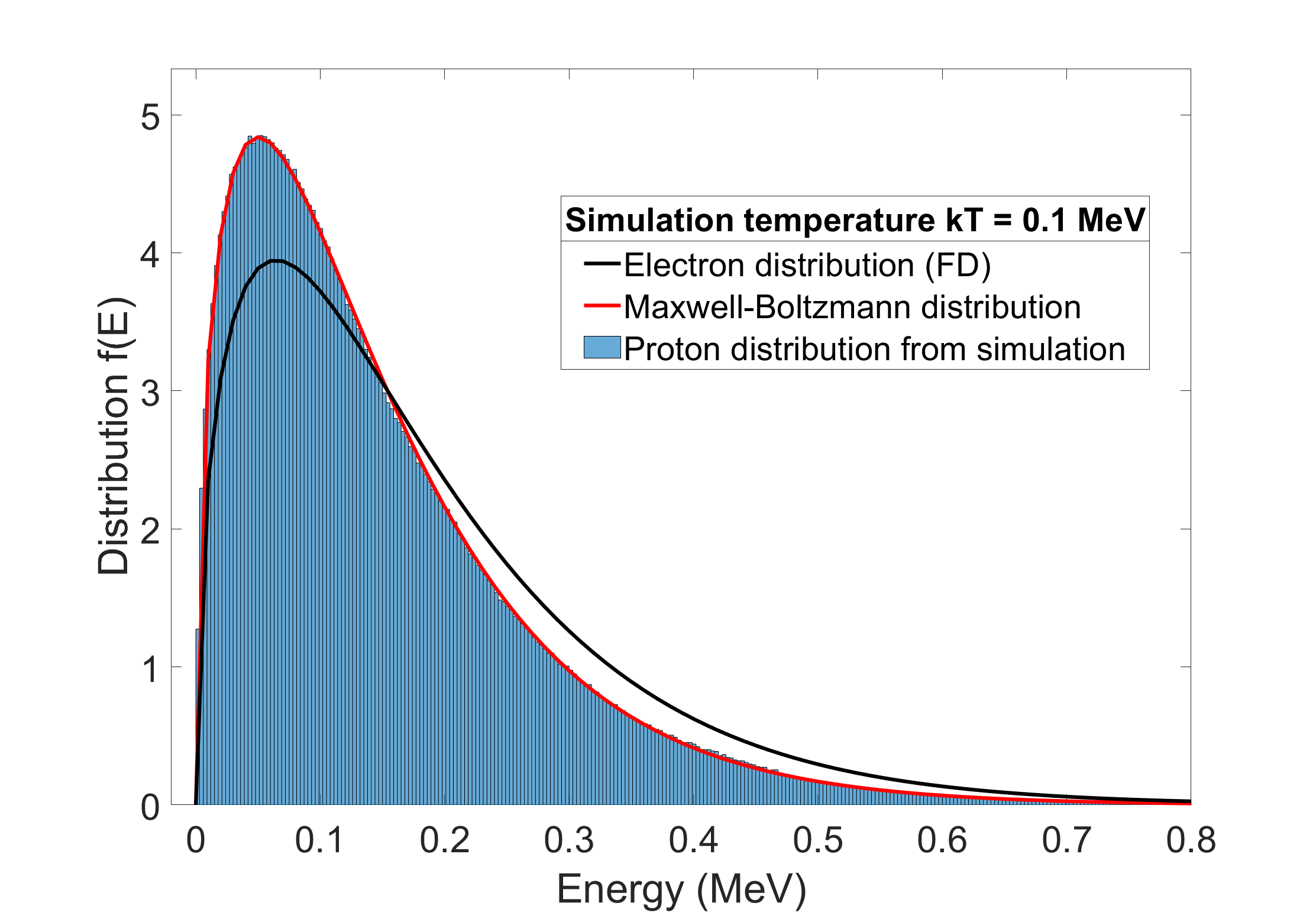}
\includegraphics[height=2.3in,width=3.5in]{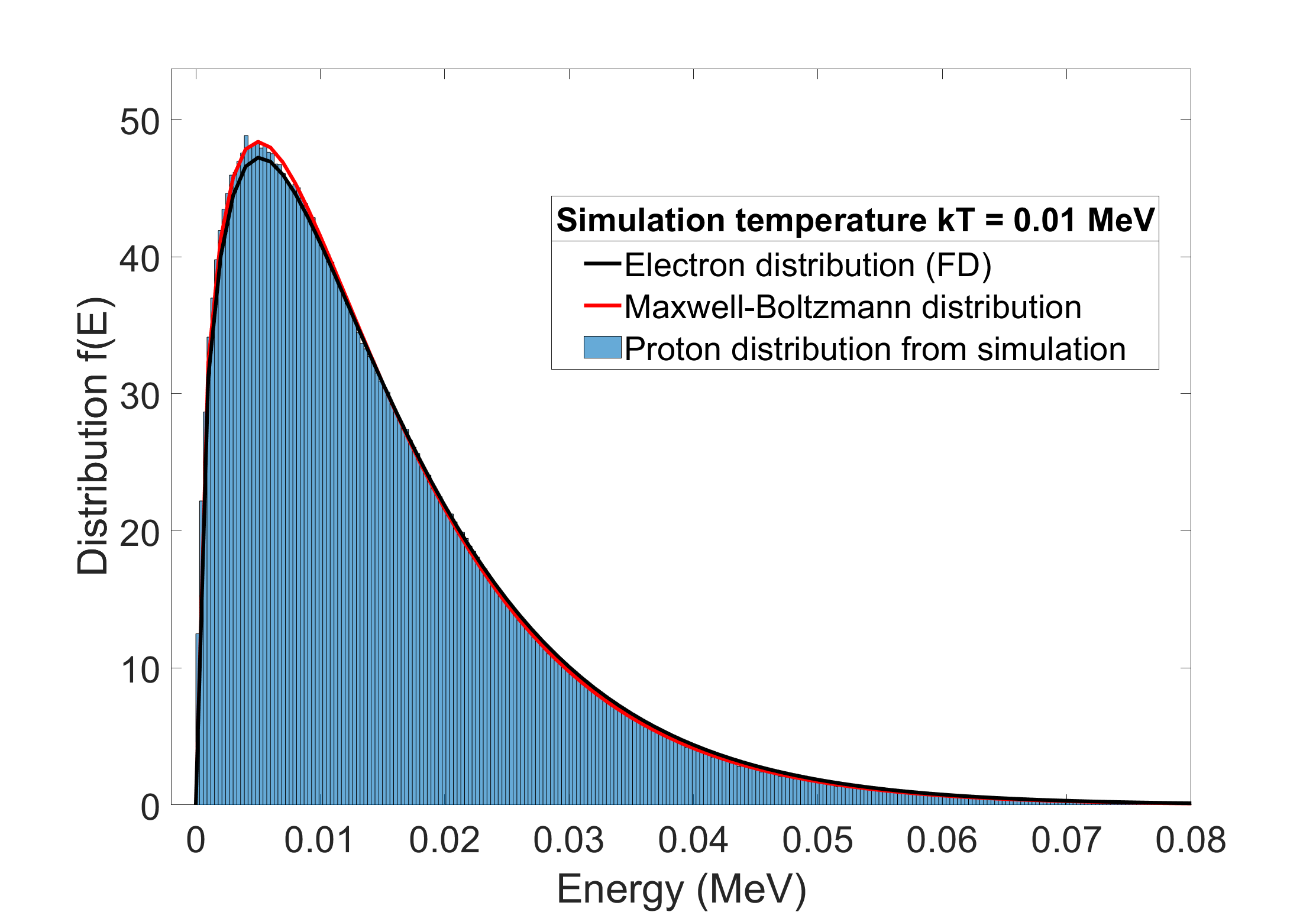}
\caption{(Color online) Monte Carlo histogram (blue bars) of the kinetic energy distribution of a nucleus scattering in a bath in 3 spatial dimensions consisting of a relativistic $e^+-e^-$ plasma (black curve) (at $kT = 1 $ MeV, $kT = 0.1 $ MeV, $kT = 0.01 $ MeV), This is compared to the kinetic energy distribution of a classical Maxwell-Boltzmann distribution (red curve).}
\label{Results:3D}
\end{figure}

Note that because the electron mass is $\sim 0.511$ MeV, the $e^+-e^-$ background is relativistic at $kT = 1$ MeV, mildly-relativistic at $kT = 0.1$ MeV and non-relativistic at $kT = 0.01$ MeV. Accordingly, in Fig. \ref{Results:3D} the $e^+-e^-$ background distributions also differ significantly from the MB distribution at $kT = 1$ MeV, and are nealy indistinguishable by $kT = 0.01$ MeV.

The equilibrium nuclear energy distribution histogram obtained as a result of the simulation is shown in Fig. \ref{Results:3D} as a blue histogram. For reference, the figures also show the FD distribution (black curve) for electrons and the MB distribution (red curve) all at the same temperature as labeled. A note should be made here that at these BBN temperatures the nuclei are dilute and non-relativistic ($kT \ll m_n c^2 \sim 939$ MeV). Hence their classical FD distribution, which is the exact distribution in absence of any other species in the surrounding, approximates to an MB distribution.

We observe from Fig. \ref{Results:3D} that at all temperatures the equilibrium thermalized proton distributions closely resemble the MB distribution corresponding to the background electron temperature. This is independent of whether or not the background electrons were relativistic. This suggests that the two species exchange energy in order to obtain the same analytical distribution, i.e. relativistic FD distribution, with the same temperature but with their respective masses for each specie. These distributions indeed indicate that, even at a common temperature, the energy partition is not the same for species with different masses. Rather, each specie attains its independent relativistic FD distribution which can, in the case of low temperature and density, approximate to an MB distribution. The simulation distributions corroborate the relativistic Boltzmann equation solution recently solved for a multicomponent gas \cite{Sasankan20,Cubero}.

In a previous work \cite{Sasankan18} we reported having observed an anomalous drift to higher energies in the nuclear energy distribution when subjected to a relativistic electron bath. The anomaly arose due to the neglect of the \textit{instantaneous viscosity} experienced by the nucleus due to it's motion w.r.t the background. Instantaneous viscosity is the effect that among electrons moving in the opposite direction and others moving in the same direction as the nucleus, the electrons moving in the opposite direction are more likely to interact with the nucleus due to their enhanced flux. This was implicitly ignored in \cite{Sasankan18} by assuming an isotropic distribution of electrons in the frame of the nucleus in step \ref{vf(v)}. In the corrected method the incoming electron is chosen based on it's flux towards the nucleus weighted by $vf(v)$. This correctly samples the electron flux according to it's velocity and direction of travel \cite{Zenitani,Melzani}.

We note that this equilibrium simulation can easily be expanded to more than one background species by adding another set of instructions on how the test particle interacts with the new species. One could then trace and study the specific interactions and dynamics of a test particle undergoing this kind of modified Brownian motion in such mixtures. 

\begin{figure}
\includegraphics[height=2.3in,width=3.5in]{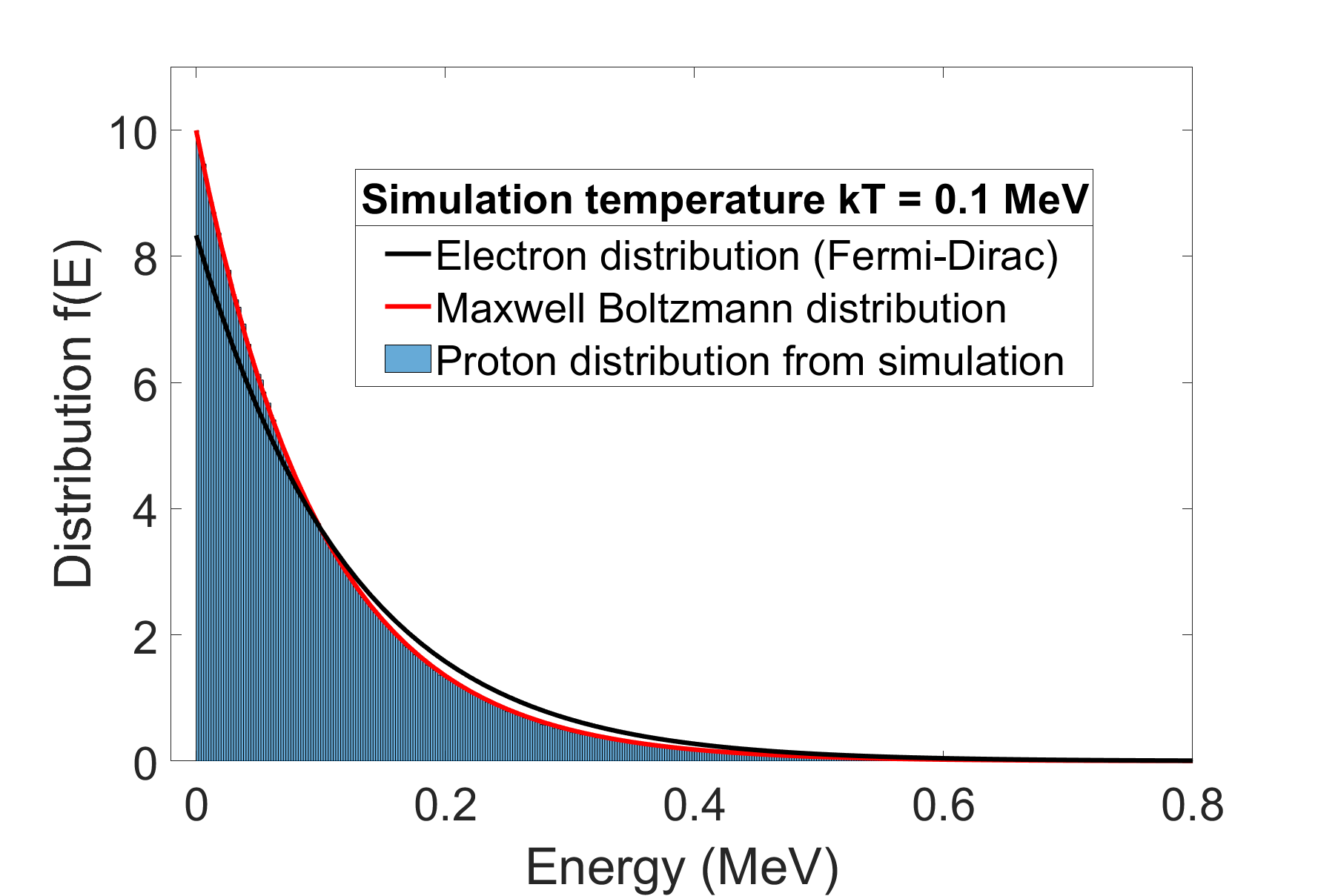}
\includegraphics[height=2.3in,width=3.5in]{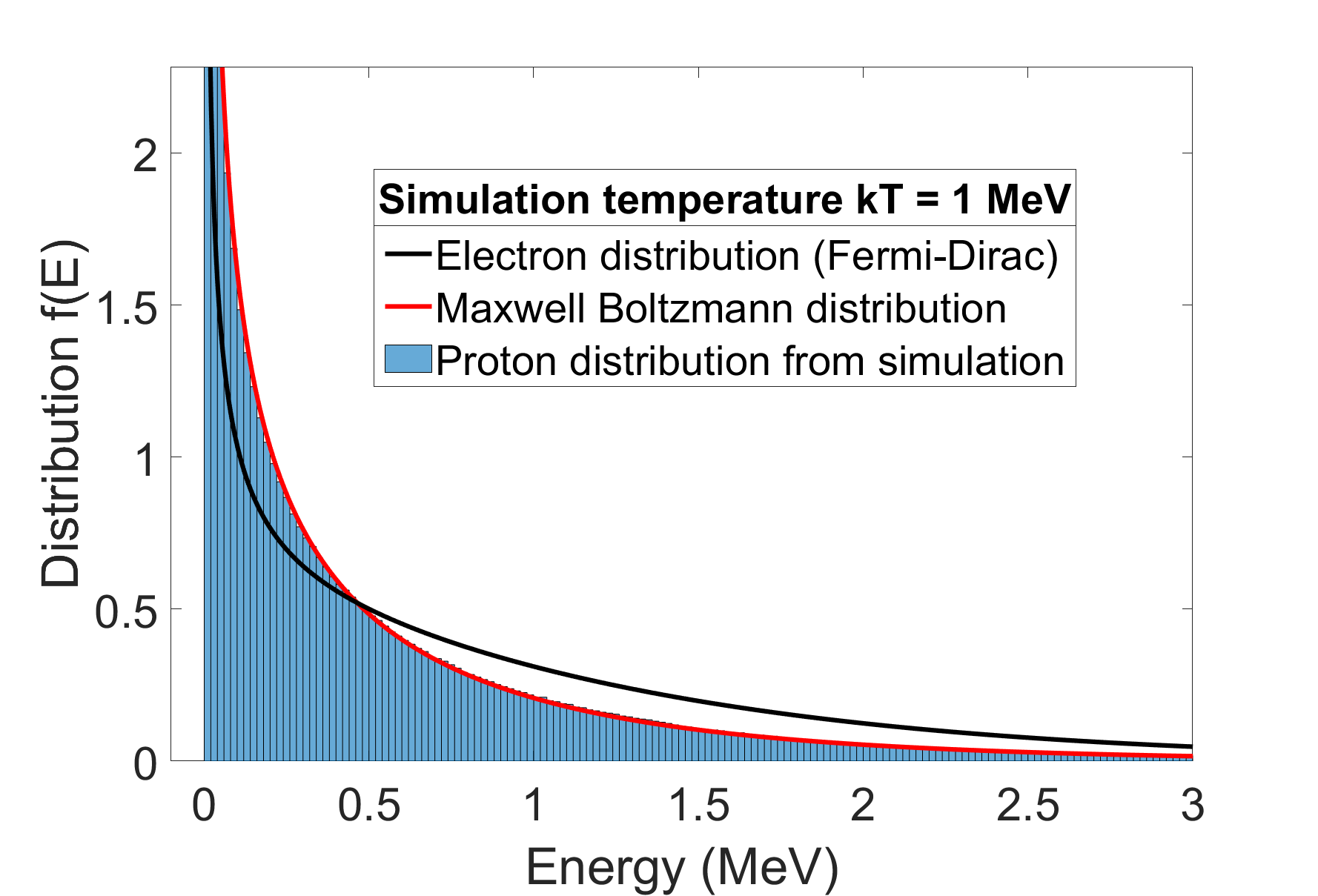}
\caption{(Color online) Monte Carlo distribution (blue bars) of the kinetic energy distribution of a nucleus scattering in baths of 2 dimensional relativistic $e^+-e^-$ plasma (black curve) (at $kT = 0.1 $ MeV) (upper panel) and 1 dimensional relativistic $e^+-e^-$ plasma (black curve) (at $kT = 1 $ MeV) (lower panel). These are compared to the kinetic energy distribution of a classical Maxwell-Boltzmann distribution (red curve).}
\label{Results:1_2D}
\end{figure}

Fig. \ref{Results:1_2D} shows the resultant distributions of the same two specie mixture from simulations performed in 2-D and 1-D for $kT = 0.1$ MeV and $kT = 1$ MeV respectively. These simulations also show the same agreement between the equilibrium nuclear energy distribution (blue histogram) and the MB distribution (red curve) for the temperature corresponding to the background electron temperature. The 1-D case has been previously studied by simulating a non-dilute mixture of two species with the resultant distribution of each specie being their respective FD distribution \cite{Cubero}. This leads to a classical MB distribution when applied to our case where the nucleus is non-relativistic as we obtained by our Monte Carlo simulation. Hence, these simulation results corroborate the result obtained previously by Cubero et al. \cite{Cubero}.

\section{Evolution toward thermalization of relativistic hadrons}
As another application we consider the injection of energetic hadrons (e.g. protons) due, for example, to the decay of a relic massive unstable particle generated during a previous epoch in the early universe. This could occur by various scenarios described, for example, in \cite{Jedamzik04,Kawasaki05,Jedamzik06, Kusakabe09, Cyburt10, Voronchev12, Kusakabe14}. As one injected particle equilibrates another is injected, so the equilibrium distribution function will then depend upon the abundance and rate of injection of energetic particles by decay.

In Ref.~\cite{Kawasaki05} for example a Monte Carlo event generator was used to calculate the spectrum of hadrons produced by the decay of a long lived exotic $X$ particle. The evolution of the hadronic shower was then studied along with its impact on the production of light nuclei in BBN. The hadron shower itself involves a complicated spectrum including relatively long-lived $\pi^\pm$ and $K^{0,\pm}$ and nucleons ($p, n, \bar p, \bar n$). At later times and lower temperatures, $kT< 0.1$ keV, mesons decay before they interact with nuclei. However, the high-energy protons and neutrons can continue to interact with the background light-elements produced during BBN. The injected spectrum of these nucleons produced during the decay was analyzed in Ref.~\cite{Kawasaki05}. It typically peaks around a kinetic energy of 10 GeV and spans a range of energies from zero up to about 20 GeV. These nucleons can scatter from and dissociate light elements produced during BBN. However, it is important to clarify the timescales for both thermalization and interaction of the energetic nucleons as they are formed by $X$-particle decay.

As an illustration of how the present code could be adapted to this application we consider the simple case of 10 GeV protons injected into the primordial plasma. That is, we follow the evolution of a proton injected with a delta-function kinetic energy of 10 GeV in a bath of electrons at $kT = 0.025$ MeV. We then follow the evolution of the distribution in time following multiple scattering.

\begin{figure*}
        \begin{subfigure}[b]{0.4\textwidth}
        \includegraphics[width=0.8\textwidth]{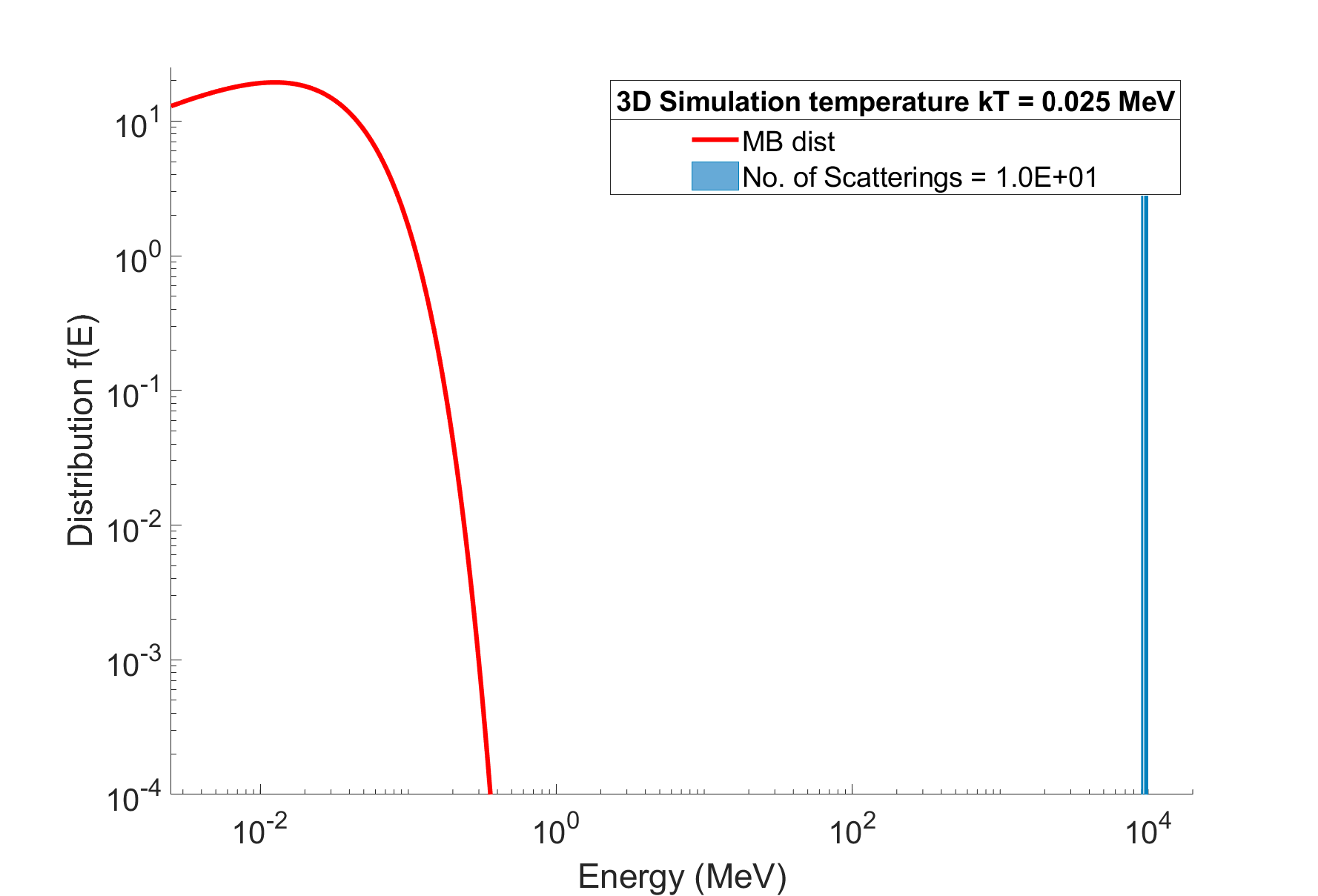}
        \caption{}
        \label{fig:therm_a}
    \end{subfigure}
    \begin{subfigure}[b]{0.4\textwidth}
        \includegraphics[width=0.8\textwidth]{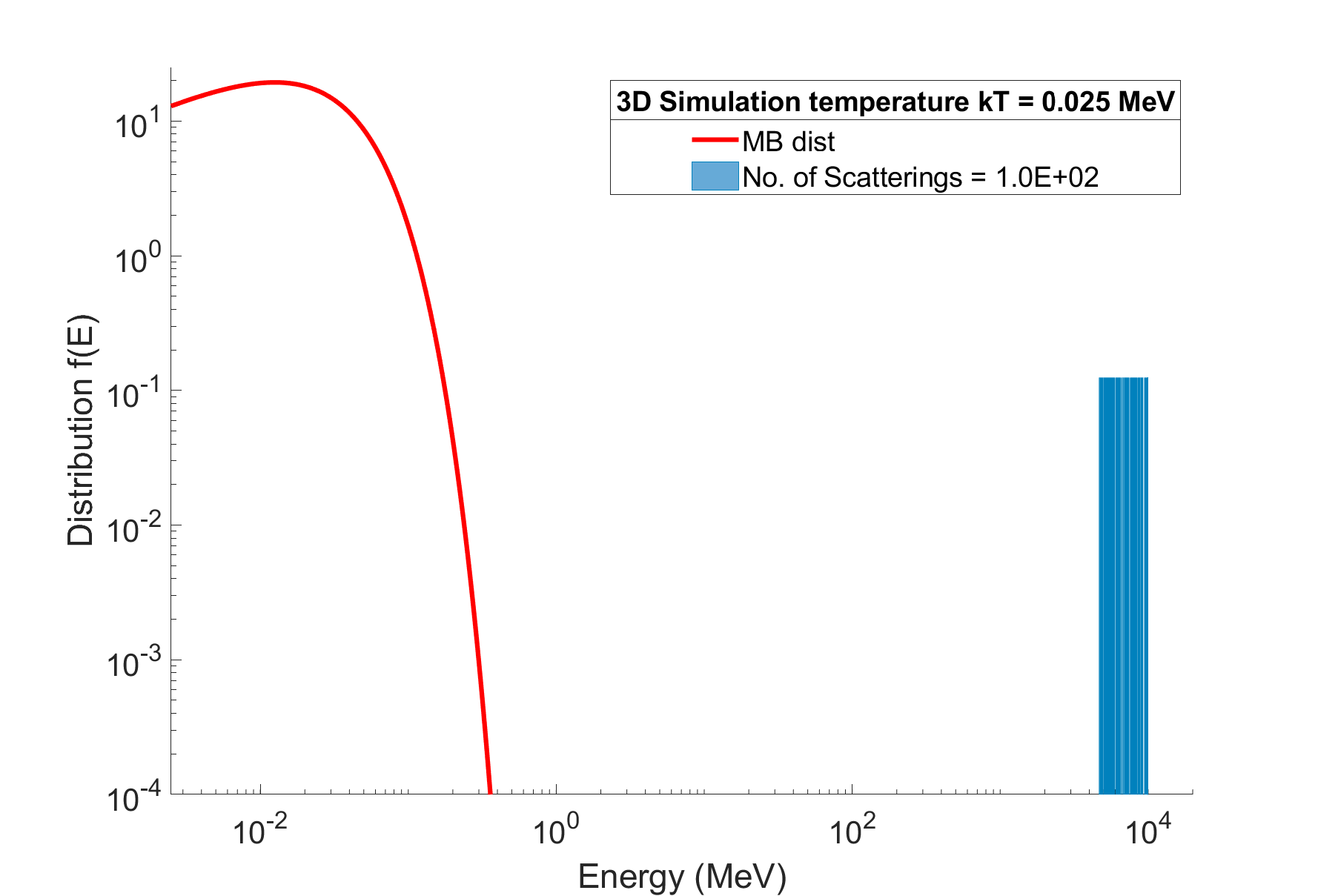}
        \caption{}
        \label{fig:therm_b}
    \end{subfigure}
    
    \begin{subfigure}[b]{0.4\textwidth}
        \includegraphics[width=0.8\textwidth]{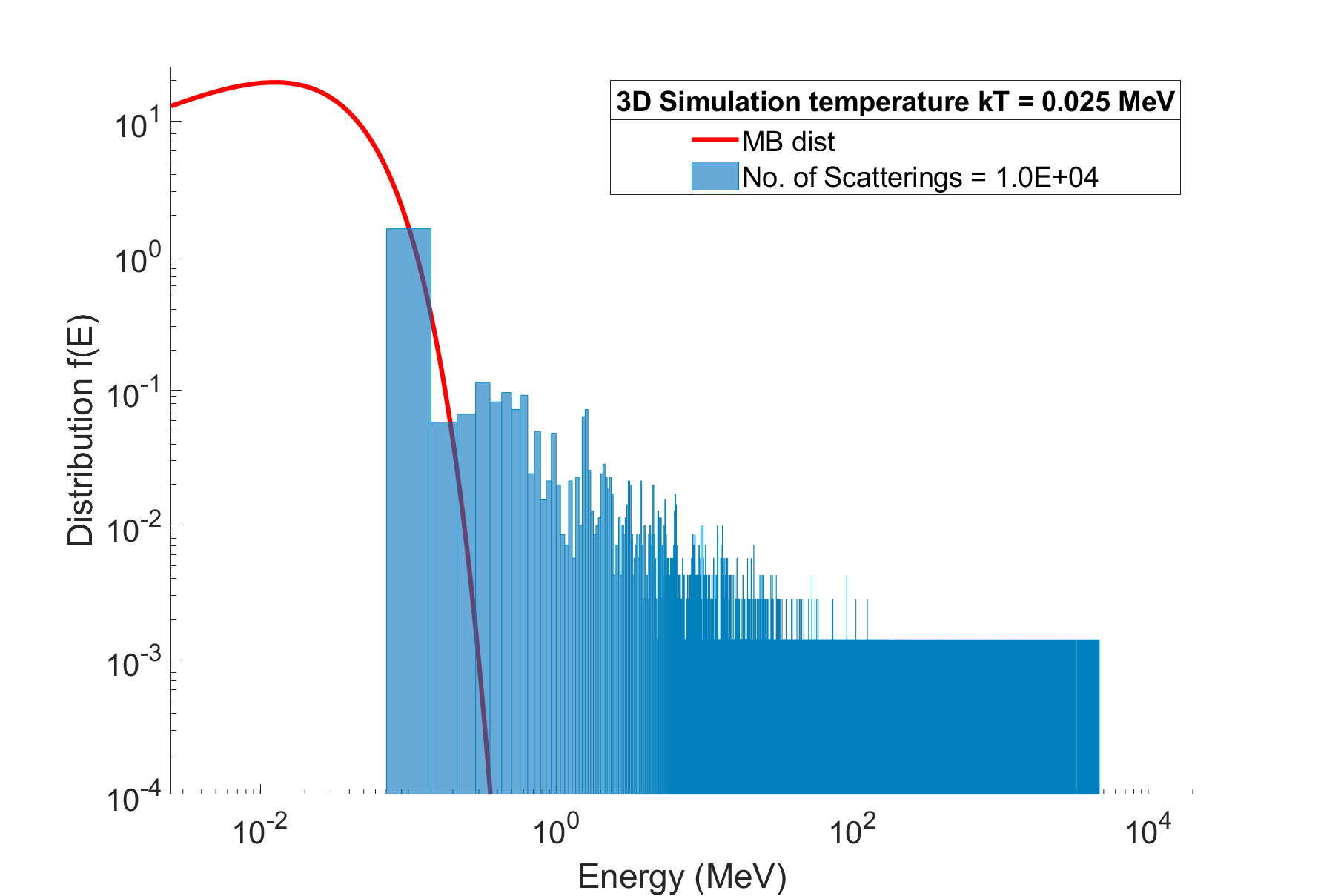}
        \caption{}
    \end{subfigure}
    \begin{subfigure}[b]{0.4\textwidth}
        \includegraphics[width=0.8\textwidth]{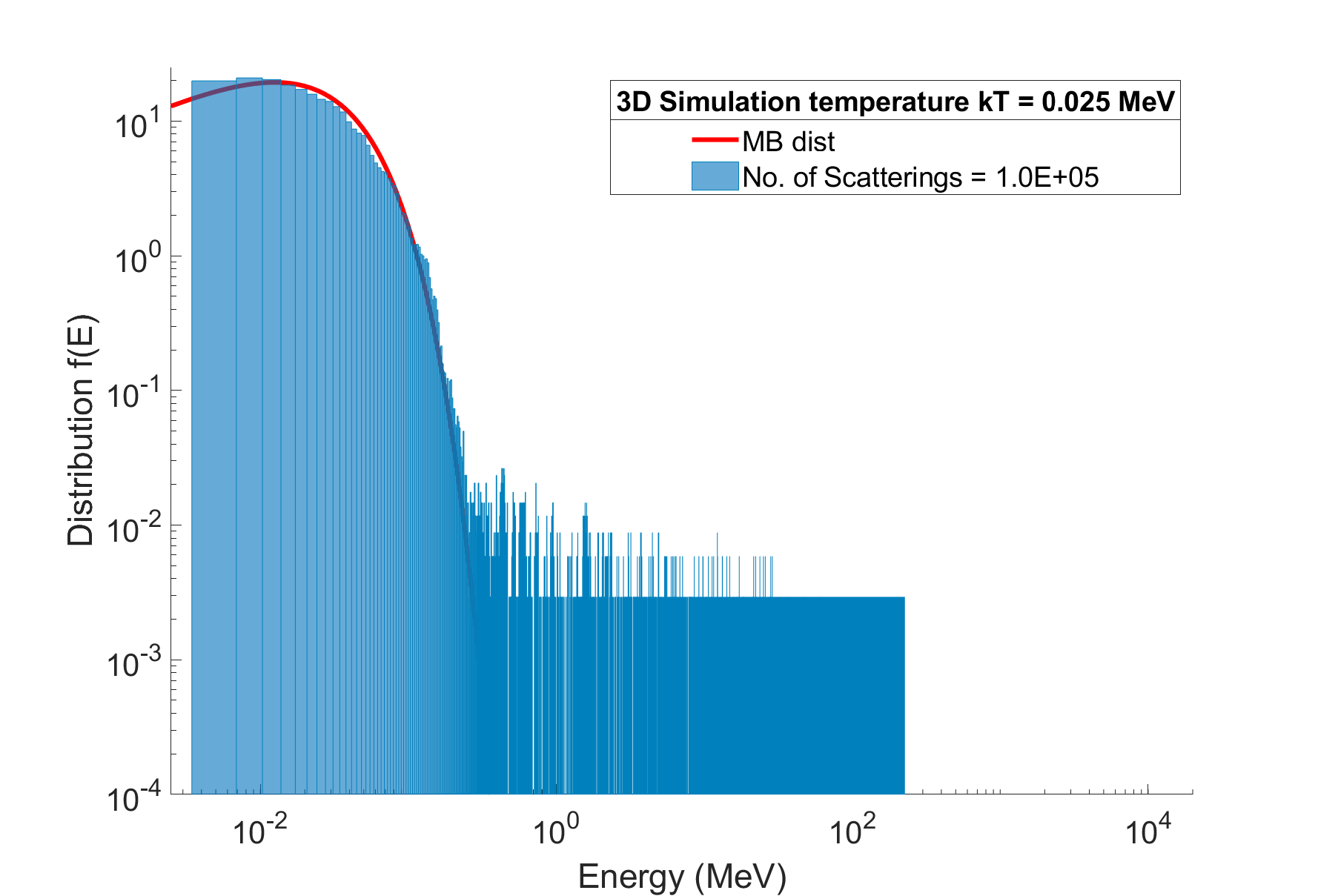}
        \caption{}
    \end{subfigure}
    
    \begin{subfigure}[b]{0.4\textwidth}
        \includegraphics[width=0.8\textwidth]{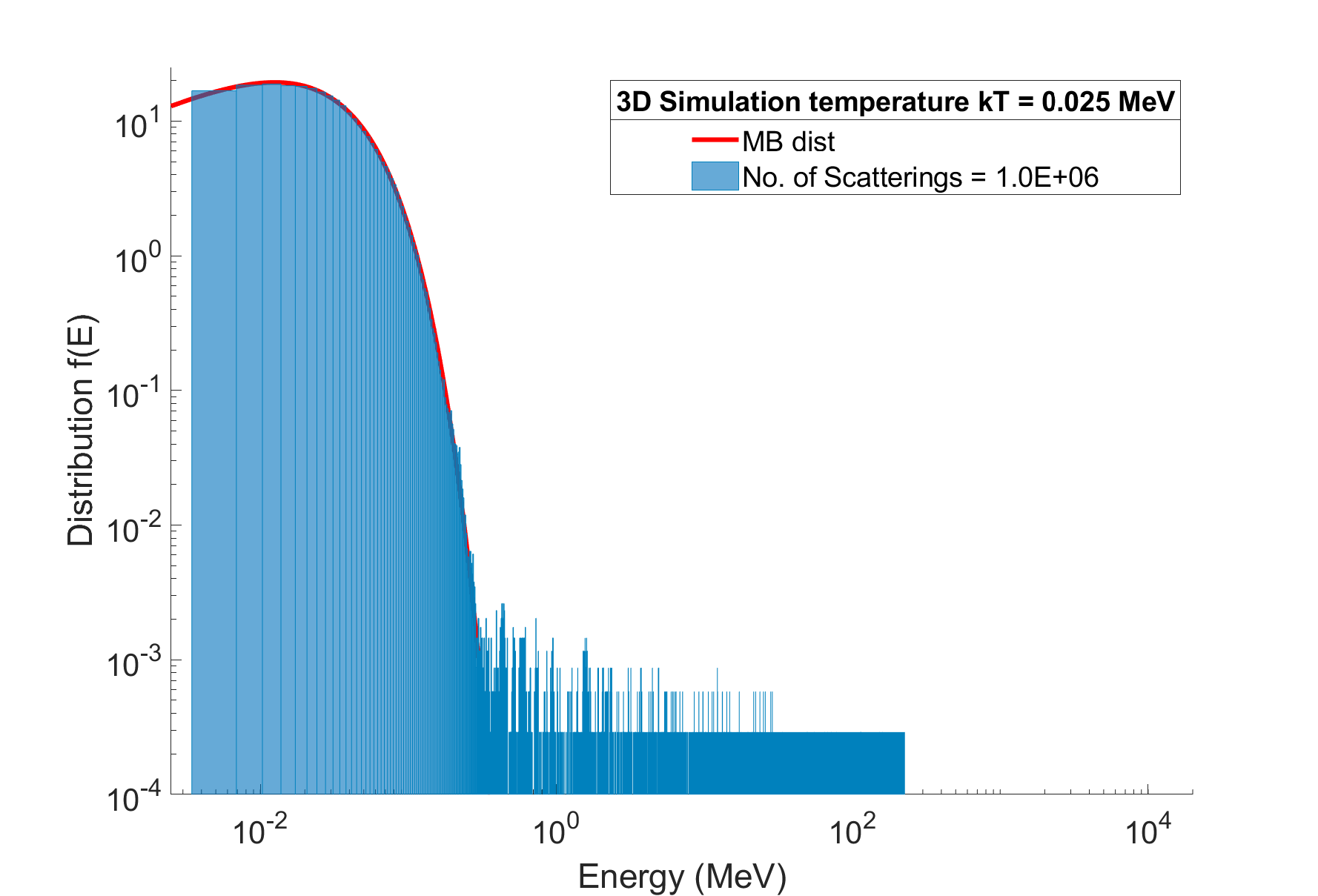}
        \caption{}
        \label{fig:therm_e}
    \end{subfigure}
    \begin{subfigure}[b]{0.4\textwidth}
        \includegraphics[width=0.8\textwidth]{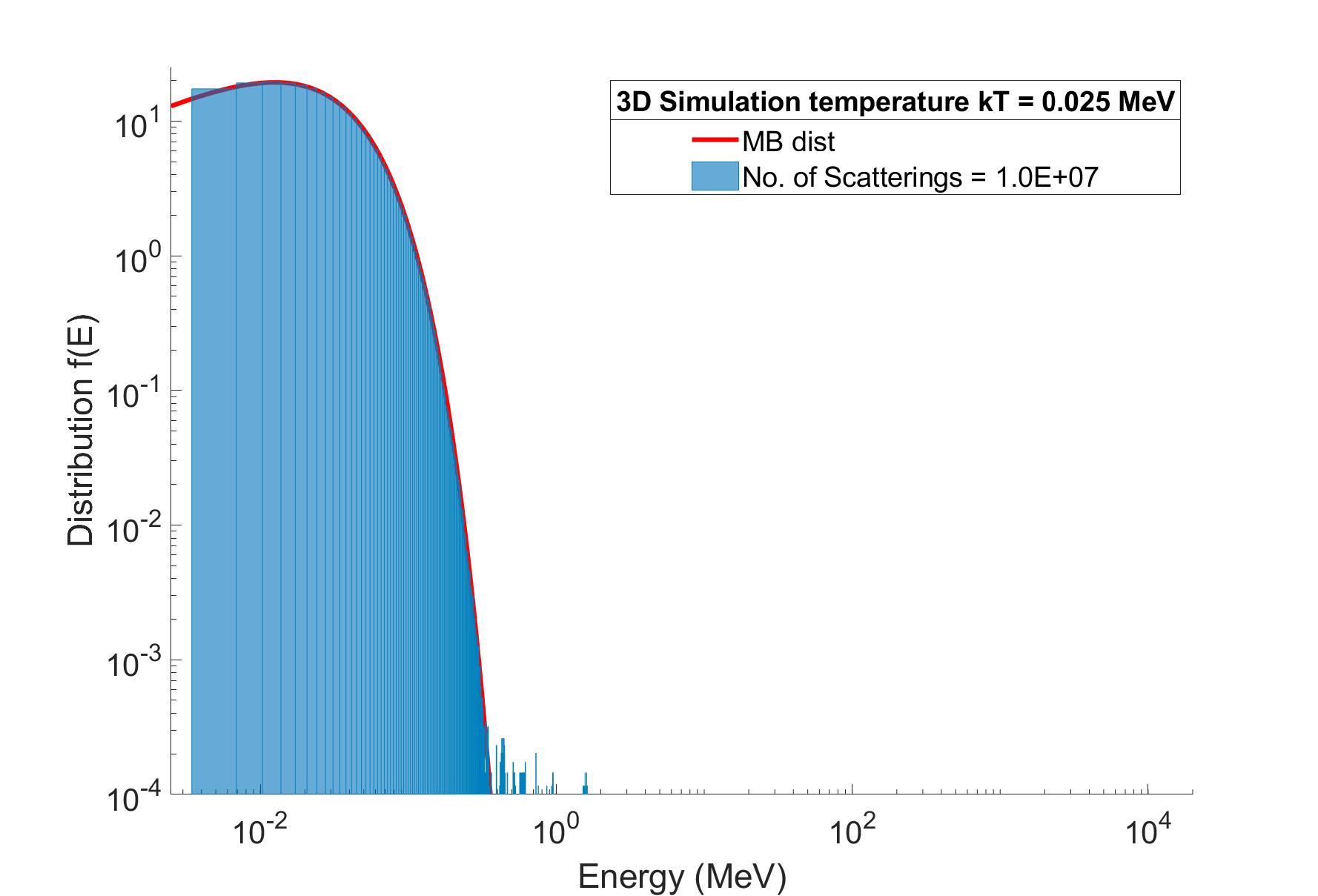}
        \caption{}
        \label{fig:therm_f}
    \end{subfigure}
\caption{The blue histogram shows the simulated progression of the normalized distribution function of protons after the indicated number of scatterings. Protons were injected with 10 GeV of kinetic energy (total energy of 10.939 GeV) and are thermalized by electron scattering in the background BBN plasma at $kT = 0.025$ MeV. The apparent rectangle at high energy in figures a-e is the result of small statistics on a log-log plot for particles in those bins. The red line shows the distribution function expected for an MB distribution at this temperature.}
\label{fig:therm}
\end{figure*}

Figure \ref{fig:therm} illustrates the evolution of the spectrum of the injected particles in a log-log plot after $10^1, 10^2, 10^4, 10^5, 10^6,~{\rm and}~ 10^7$ scatterings at a background temperature of kT = 0.025 MeV. This is compared with the expected thermalized nucleon distribution, i.e. an MB distribution (red curve). The spectrum of the proton starts as a delta function at 10 GeV, and can be seen in Fig. \ref{fig:therm_a} as a blue bar. The proton starts losing energy to it's surrounding as background electrons interact as seen in the distributions in Figs. \ref{fig:therm_b}- \ref{fig:therm_e}. After having scattered with the background enough times the protons assume a MB distribution as can be seen in Fig. \ref{fig:therm_f}.


The average scattering rate per nucleon $\Gamma$ is approximately given by $\Gamma = n_e \sigma v$, where $n_e$ is the electron density, $\sigma$ the scattering cross section, assumed to be the 10 times the size of a nucleus (to be consistent with our cutoff of impact parameters stated in Step 8 of our algorithm), and $v$ is the average relative thermal velocity. So, for an average background electron density of $\sim 10^{20}$ cm$^{-3}$ at kT= 0.025 MeV, the many elastic collisions shown in Fig. \ref{fig:therm_a}-\ref{fig:therm_f} i.e. $10^1, 10^2, 10^4, 10^5, 10^6, 10^7$ scatterings approximately corresponds to times of order 35 $\mu$s, 0.35 ms, 35 ms, 0.35 s, 3.5 s, ~{\rm and}~ 35 seconds during the big bang. For that temperature of the universe as the universe cools down and changes, the cosmic time changes by less than the 35s needed for the equilibrium to be achieved. Thus, the remnant tail at high energy between $10^6$ and $10^7$ scattering events, may remain and impact nuclear reaction rates during BBN. This will be explored in a future work. Note that for high-energy protons, such as those with 10 GeV considered here, the most important energy loss process is not via Coulomb scattering from electrons, but from strong interaction scattering with background nucleons \cite{Meyer, Boyd}. In this case the equilibration times estimated here are upper limits.

\section{Conclusion}
We have presented a Monte Carlo algorithm for the simulation of multidimensional, multicomponent relativistic thermalization. This method could be used for simulating a bath of multiple different species to replicate environmental conditions any one test particle experiences. 

We illustrated two applications of this algorithm for the solution of the distribution function for a heavy particle experiencing Brownian-like scattering in a bath of relativistic light particles. The test conditions were motivated from big bang nucleosynthesis, as charged nuclei interact with surrounding relativistic constituents, i.e. electrons and positrons. The temperature range we chose was between 0.01 MeV to 1 MeV appropriate to BBN.

Our first test simulation of the equilibrium thermalized distribution functions at various temperatures corroborates the expected results, i.e. the proton distribution is found to be very close to the MB distribution. To our knowledge this is the first fully relativistic multicomponent simulation in three spatial dimensions of such relativistic Brownian motion.

As a second test we have evaluated the thermalization of energetic hadrons injected into a background $e^+-e^-$ plasma at a temperature of 0.025 MeV. This illustrates how the nuclear spectrum may be distorted due to a continuous injection of nonthermal particles during the big bang.

\section{Acknowledgment}
Work at the University of Notre Dame is supported by the U.S. Department of Energy under Nuclear Theory Grant No. DE-FG02-95-ER40934. One of the authors (M.K.) acknowledges support from the Japan Society for the Promotion of Science (27.957).

\appendix

\section{Derivation of Lorentz transformation of $f(\mathbf{v})$ to $f'(\mathbf{v'})$}~~.

For our selection of the colliding background particle we needed the background particle distribution in the rest frame of the test particle. Such a distribution would have to be obtained by performing a transformation from the distribution in the cosmic frame. The difficulty in finding a Lorentz-invariant distribution that also satisfies simulation results has been discussed previously \cite{Curado}. Here we derive the Lorentz transformed distribution of a relativistic gas in a moving frame. We start with relativistic distributions, i.e. relativistic FD or its non-degenerate approximation Maxwell-J\"uttner distribution in the rest frame, and find the equivalent distribution in the moving frame.

We begin with the conservation relation regarding the distribution functions \cite{Zenitani,Landau-Lifshitz},
\begin{equation}\label{eq:1}
     f'(\mathbf{x'},\mathbf{u'}) = f(\mathbf{x},\mathbf{u})~~,
\end{equation}
where the prime ($'$) denotes quantities in the moving frame and the unprimed quantities are in lab frame, i.e. the frame at rest w.r.t the background fluid. $\mathbf{x}$ are the spatial coordinates and $\mathbf{u} = \gamma \mathbf{v} = \frac{\mathbf{p}}{m}$ are spatial parts of the four velocity. Using this we want to find $f'(\mathbf{v'})$, and we know $f(\textbf{v})$ and $f(\textbf{u})$ are a relativistic FD distribution and a Maxwell-J\"uttner distribution for the two cases, for electrons as they are in the background fluid's rest frame.

First, solving for $f'(\textbf{u'})$ using Eq.~(\ref{eq:1}),
\begin{align}
     f'(\mathbf{x'},\mathbf{u'}) d^3\mathbf{x'}&= f(\mathbf{x},\mathbf{u}) d^3\mathbf{x'} \\
     \int f'(\mathbf{x'},\mathbf{u'}) d^3\mathbf{x'} &= \int f(\mathbf{x},\mathbf{u}) d^3\mathbf{x} \times \frac{d^3\mathbf{x'}}{d^3\mathbf{x}}.
\end{align}
Note that here and in the rest of the appendix $\times$ indicates a simple multiplication of two scalars and not a cross-product. We know, in our case $f(\mathbf{x},\mathbf{u})$ and $f'(\mathbf{x'},\mathbf{u'})$ are position independent, i.e. we expect the distribution to be independent of coordinates, but only different in different reference frames. Hence, the integration simply gives the volume in the two frames, albeit contracted by the relevant Lorentz factors. Therefore,
\begin{align}
     \implies f'(\mathbf{u'}) = f(\mathbf{u}) \times \frac{\gamma_o}{\gamma'_V}~~,
     \label{eq:5}
\end{align}
where $\gamma'_V$ and $\gamma_o$ are the Lorentz factors for the speed of the two frames, i.e. they should be $\gamma_o = 1$ (for the lab frame, which is at rest with respect to the gas cloud) and $\gamma'_V = \frac{1}{\sqrt{1-V^2}}$ (where $V$ is the speed of the moving frame).

\subsection{1 Dimension}
Now since we want $f'(\mathbf{v'})$, we multiply Eq.~(\ref{eq:5}) by $d\mathbf{u'}$ to get:
\begin{equation}
     f'(\mathbf{u'})d\mathbf{u'} = f(\mathbf{u}) \times \frac{\gamma_o}{\gamma'_V}d\mathbf{u'}~~.
\end{equation}
But we know that in 1-D, the change of variable from $\mathbf{u'}$ to $\mathbf{v'}$ is as:
\begin{equation}
    f'(\mathbf{u'})d\mathbf{u'} = f'(\mathbf{v'})d\mathbf{v'}~~.
\end{equation}
Therefore, by combining the last two equations, we have
\begin{align}
     f'(\mathbf{v'})d\mathbf{v'}&= f(\mathbf{u}) \times \frac{\gamma_o}{\gamma'_V}d\mathbf{u'} \\
     \implies f'(\mathbf{v'}) &= f(\mathbf{u}) \times \frac{\gamma_o}{\gamma'_V}\frac{d\mathbf{u'}}{d\mathbf{v'}}~~.
\end{align}
Since,
\begin{align}
    \mathbf{u'} &= \gamma'\mathbf{v'} \\
    \implies \mathbf{u'} &= \frac{1}{\sqrt{1-v'^2}}\mathbf{v'} \\
    \implies d\mathbf{u'} &= \gamma'^3d\mathbf{v'}~~,
\end{align}
we have,
\begin{equation}
    f'(\mathbf{v'}) = \frac{\gamma_o}{\gamma'_V} f(\mathbf{u}) \gamma'^3 ~~,
    \label{eq:fv1}
\end{equation}
and we know:
\begin{align}
    f_{FD,1D}(\mathbf{u}) &= (\frac{1}{n}) \frac{1}{(1+\exp(\frac{\gamma mc^2}{kT}))} \\
    f_{MJ,1D}(\mathbf{u}) &= \frac{\exp(-\frac{\gamma mc^2}{kT})}{2mcK_1\Big(\frac{mc^2}{kT}\Big)}~~,
\end{align}
where, ${1}/{n}$ is an approximate normalization constant. The factors independent of $\gamma$ are irrelevant for our purpose as they are independent of $\mathbf{v}$ and $\mathbf{u}$. Plugging $f_{FD,1D}(\mathbf{u})$ and $f_{MJ,1D}(\mathbf{u})$ in Eq.~(\ref{eq:fv1}) gives,
\begin{align}
     f_{FD,1D}'(\mathbf{v'}) &= (\frac{1}{n}) \frac{\gamma_o}{\gamma'_V} \gamma'^3 \frac{1}{(1+\exp(\frac{\gamma mc^2}{kT}))} \\
     f_{MJ,1D}'(\mathbf{v'}) &= \frac{\gamma_o}{\gamma'_V} \gamma'^3 \frac{\exp(-\frac{\gamma mc^2}{kT})}{2mcK_1\Big(\frac{mc^2}{kT}\Big)}~~.
\end{align}
Substituting $\gamma = \gamma'\gamma_V(1+Vv')$ from \cite{Zenitani} gives,
\begin{align}
     f_{FD,1D}'(\mathbf{v'}) &= (\frac{1}{n})\frac{\gamma_o}{\gamma'_V} \gamma'^3 \frac{1}{(1+\exp(\frac{\gamma'\gamma_V(1+Vv')mc^2}{kT}))} \\
     f_{MJ,1D}'(\mathbf{v'}) &= \frac{\gamma_o}{\gamma'_V} \gamma'^3 \frac{\exp(-\frac{\gamma'\gamma_V(1+Vv')mc^2}{kT})}{2mcK_1\Big(\frac{mc^2}{kT}\Big)}~~.
\end{align}
This is the needed Lorentz-transformed distribution. This distribution agrees with the previous study that obtained this distribution by building a 1-D thermalization simulation with all the background particles tracked \cite{DunkelHanggi}. In our 1-D simulation we use $|v'|f'(\mathbf{v'})$ for sampling the velocity $\mathbf{v'}$ at which electrons come to scatter from the nucleus.

\subsection{2 Dimensions}
For the 2-D and 3-D cases we employ the same analytical procedure as was outlined in the 1-D case to obtain the 2-D and 3-D Lorentz transformed velocity distribution.

Since we want $f'(\mathbf{v'})$, multiply Eq(\ref{eq:5}) by $d^2\mathbf{u'}$ to get:
\begin{equation}
     \implies f'(\mathbf{u'})d^2\mathbf{u'} = f(\mathbf{u}) \times \frac{\gamma_o}{\gamma'_V}d^2\mathbf{u'}~~.
\end{equation}
However, we know that in 2-D, the change of variable from $\mathbf{u'}$ to $\mathbf{v'}$ is as:
\begin{equation}
    f'(\mathbf{u'})d^2\mathbf{u'} = f'(\mathbf{v'})d^2\mathbf{v'}~~.
\end{equation}
Therefore, by combining the last two equations,
\begin{align}
     f'(\mathbf{v'})d^2\mathbf{v'}&= f(\mathbf{u}) \times \frac{\gamma_o}{\gamma'_V}d^2\mathbf{u'} \\
     \implies f'(\mathbf{v'}) &= f(\mathbf{u}) \times \frac{\gamma_o}{\gamma'_V}\frac{d^2\mathbf{u'}}{d^2\mathbf{v'}}~~.
\end{align}
To find $\frac{d^2\mathbf{u'}}{d^2\mathbf{v'}}$ we need to find the Jacobian matrix
\begin{equation}
J = 
    \begin{bmatrix}
    \frac{\partial u_x}{\partial v_x} & \frac{\partial u_x}{\partial v_y} \\
    \frac{\partial u_y}{\partial v_x} & \frac{\partial u_y}{\partial v_y}
    \end{bmatrix}~~.
\end{equation}
The change in the volume element in the change of space of integration is given by the determinant of the Jacobian $|J|$, i.e.
\begin{equation}
    \frac{d^2\mathbf{u'}}{d^2\mathbf{v'}} = |J'| = \gamma'^4~~.
\end{equation}

Therefore, we have:
\begin{equation}
    f'(\mathbf{v'}) = \frac{\gamma_o}{\gamma'_V} f(\mathbf{u}) \gamma'^4~~,
    \label{eq:fv2}
\end{equation}
and we know,
\begin{align}
    f_{FD,2D}(\mathbf{u}) &= (\frac{1}{n})\frac{1}{\Big(1+\exp\Big(\frac{\gamma mc^2}{kT}\Big)\Big)}~~, \\
    f_{MJ,2D}(\mathbf{u}) &= \frac{c^2m^2}{2\pi kT(mc^2+kT)}\exp\Big(-\frac{(\gamma-1)mc^2}{kT}\Big)~~,
\end{align}
where, ${1}/{n}$ is the appropriate normalization constant. Plugging $f_{FD,2D}(\mathbf{u})$ and $f_{MJ,2D}(\mathbf{u})$ in Eq.~(\ref{eq:fv2}) gives,
\begin{align}
     f_{FD,2D}'(\mathbf{v'}) &= (\frac{1}{n})\frac{\gamma_o}{\gamma'_V} \gamma'^4\frac{1}{\Big(1+\exp\Big(\frac{\gamma mc^2}{kT}\Big)\Big)}~~, \\
     f_{MJ,2D}'(\mathbf{v'}) &= \frac{\gamma_o}{\gamma'_V} \gamma'^4\frac{c^2m^2}{2\pi kT(mc^2+kT)}\exp\Big(-\frac{(\gamma-1)mc^2}{kT}\Big)~~.
\end{align}

Then substituting $\gamma = \gamma'\gamma_V(1+Vv_x')$ from \cite{Zenitani} gives,
\begin{eqnarray}
     f_{FD,2D}'(\mathbf{v'}) &=& (\frac{1}{n})\frac{\gamma_o}{\gamma'_V} \gamma'^4\frac{1}{\Big(1+\exp\Big(\frac{\gamma'\gamma_V(1+Vv_x')mc^2}{kT}\Big)\Big)}~~, \\
     f_{MJ,2D}'(\mathbf{v'}) &=& \frac{\gamma_o}{\gamma'_V} \gamma'^4 \frac{c^2m^2}{2\pi kT(mc^2+kT)} \nonumber \\
     && \times\exp\Big(-\frac{(\gamma'\gamma_V(1+Vv_x')-1)mc^2}{kT}\Big)~~.
\end{eqnarray}

This distribution is a new result. In our 2-D simulation we use $\sigma (v')|v'|f'(\mathbf{v'})$ for sampling the velocity $\mathbf{v'}$ at which electrons scatter from the nucleus.

\subsection{Three dimensions}
Since we want $f'(\mathbf{v'})$, we multiply Eq.~(\ref{eq:5}) by $d^3\mathbf{u'}$ to get:
\begin{equation}
     \implies f'(\mathbf{u'})d^3\mathbf{u'} = f(\mathbf{u}) \times \frac{\gamma_o}{\gamma'_V}d^3\mathbf{u'}
\end{equation}
But we know in 3-D, change of variable from $\mathbf{u'}$ to $\mathbf{v'}$ is:
\begin{equation}
    f'(\mathbf{u'})d^3\mathbf{u'} = f'(\mathbf{v'})d^3\mathbf{v'}~~.
\end{equation}
Therefore, by combining the last two equations,
\begin{align}
     f'(\mathbf{v'})d^3\mathbf{v'}&= f(\mathbf{u}) \times \frac{\gamma_o}{\gamma'_V}d^3\mathbf{u'} \\
     \implies f'(\mathbf{v'}) &= f(\mathbf{u}) \times \frac{\gamma_o}{\gamma'_V}\frac{d^3\mathbf{u'}}{d^3\mathbf{v'}}~~.
\end{align}
To find $\frac{d^3\mathbf{u'}}{d^3\mathbf{v'}}$ we again need to find the Jacobian matrix
\begin{equation}
J = 
    \begin{bmatrix}
    \frac{\partial u_x}{\partial v_x} & \frac{\partial u_x}{\partial v_y} & \frac{\partial u_x}{\partial v_z}\\
    \frac{\partial u_y}{\partial v_x} & \frac{\partial u_y}{\partial v_y} & \frac{\partial u_y}{\partial v_z}\\
    \frac{\partial u_z}{\partial v_x} & \frac{\partial u_z}{\partial v_y} & \frac{\partial u_z}{\partial v_z}
    \end{bmatrix}~~.
\end{equation}
The change in the volume element in the change of space of integration is given by the determinant of the Jacobian $|J|$. Thus,
\begin{equation}
    \frac{d^3\mathbf{u'}}{d^3\mathbf{v'}} = |J'| = \gamma'^5~~.
\end{equation}

So that,
\begin{equation}
    f'(\mathbf{v'}) = \frac{\gamma_o}{\gamma'_V} f(\mathbf{u}) \gamma'^5~~,
    \label{eq:fv3}
\end{equation}
and we know,
\begin{align}
    f_{FD,3D}(\mathbf{u}) &= \bigg(\frac{1}{n}\bigg) \frac{1}{\Big(1+\exp\Big(\frac{\gamma mc^2}{kT}\Big)\Big)}~~, \\
    f_{MJ,3D}(\mathbf{u}) &= \frac{m}{4\pi ckTK_2\Big(\frac{mc^2}{kT}\Big)}\exp\Big(-\frac{\gamma mc^2}{kT}\Big)~~,
\end{align}
where, ${1}{n}$ is the approximate normalization constant. As noted above ehe constants independent of $\gamma$ are irrelevant for our purpose as they are independent of $\mathbf{v}$ and $\mathbf{u}$. Plugging $f_{FD,3D}(\mathbf{u})$ and $f_{MJ,3D}(\mathbf{u})$ in Eq.~(\ref{eq:fv3}) gives,
\begin{align}
     f_{FD,3D}'(\mathbf{v'}) &= (\frac{1}{n})\frac{\gamma_o}{\gamma'_V} \gamma'^5 \frac{1}{\Big(1+\exp\Big(\frac{\gamma mc^2}{kT}\Big)\Big)}~~, \\
     f_{MJ,3D}'(\mathbf{v'}) &= \frac{\gamma_o}{\gamma'_V} \gamma'^5 \frac{m}{4\pi ckTK_2\Big(\frac{mc^2}{kT}\Big)}\exp\Big(-\frac{\gamma mc^2}{kT}\Big)~~.
\end{align}
substituting $\gamma = \gamma'\gamma_V(1+Vv_x')$ from \cite{Zenitani} gives,
\begin{eqnarray}
     f_{FD,3D}'(\mathbf{v'}) &=& \bigg(\frac{1}{n}\bigg)\frac{\gamma_o}{\gamma'_V} \gamma'^5 \frac{1}{\Big(1+\exp\Big(\frac{\gamma'\gamma_V(1+Vv_x')mc^2}{kT}\Big)\Big)} ~~,\nonumber\\
     \\
     f_{MJ,3D}'(\mathbf{v'}) &=& \frac{\gamma_o}{\gamma'_V} \gamma'^5 \frac{m}{4\pi ckTK_2\Big(\frac{mc^2}{kT}\Big)} \nonumber \\
     &&\times \exp\Big(-\frac{\gamma'\gamma_V(1+Vv_x')mc^2}{kT}\Big)~~.
\end{eqnarray}

This distribution is a new result we found. In our 3-D simulations we use $\sigma (v')|v'|f'(\mathbf{v'})$ for sampling the velocity $\mathbf{v'}$ at which electrons scatter from the nucleus. The resultant distribution obtained for the nucleus corroborates with analytical solutions \cite{Sasankan20} and is hence tested via simulation.


\begin{references}

\bibitem{1+3} J. Dunkel, and P. H\"anggi, Phys. Rev. E 72.3 (2005): 036106.
\bibitem{1+1} J. Dunkel, and P. H\"anggi, Phys. Rev. E 71.1 (2005): 016124.
\bibitem{Cubero} D. Cubero, J. Casado-Pascual, J. Dunkel, P. Talkner, and P. H\"anggi, Phys. Rev. Lett. 99.17 (2007): 170601.
\bibitem{DunkelHanggi} J. Dunkel, and P. H\"anggi, Physics Reports 471.1 (2009): 1-73.
\bibitem{Kremer} G. M. Kremer, and W. Marques Jr, Phys. of Fluids 25.1 (2013): 017102.
\bibitem{Cercignani} C. Cercignani, G. M. Kremer, \textit{Relativistic Boltzmann Equation. In: The Relativistic Boltzmann Equation: Theory and Applications.} Progress in Mathematical Physics, Vol 22. Birkhäuser, Basel (2002).
\bibitem{Sasankan20} N. Sasankan, A. Kedia, M. Kusakabe, and G. J. Mathews, Phys. Rev. D 101, 123532 (2020).
\bibitem{Acosta} G. Cha\'con-Acosta, Guillermo, and G. M. Kremer, Phys. Rev. E 76.2, 021201 (2007).
\bibitem{github} \href{https://github.com/AtulKedia93/Multicomponent_relativistic_thermlization}{\url{https://github.com/AtulKedia93/Multicomponent_relativistic_thermlization}}
\bibitem{Tsallis} C. Tsallis, 1988, JSP, 52, 479
\bibitem{Kusakabe19} M. Kusakabe, T. Kajino, G. J. Mathews and Y. Luo, Phys. Rev. D 99, 043505 (2019).
\bibitem{Bertulani13} C. A. Bertulani, J. Fuqua, and M. S. Hussein, Astrophys. J. {\bf767}, 67, (2013).
\bibitem{Hou17} S. Q. Hou, J. J. He, A. Parikh, D. Kahl, C. A. Bertulani, T. Kajino, G. J. Mathews, and G. Zhao, Astrophys. J. {\bf 834} 165 (2017).
\bibitem{Luo19} Y. Luo, T. Kajino, M. Kusakabe, G. J. Mathews, Astrophys. J., 872, 172 (2019).
\bibitem{Jang18} D. Jang, Y. Kwon, K. Kwak, and M.-K. Cheoun, arXiv:1812.09472.
\bibitem{Jedamzik04} K. Jedamzik, Phys. Rev. D 70, 063524 (2004).
\bibitem{Kawasaki05} M. Kawasaki, K. Kohri, and T. Moroi, Phys. Rev. D 71, 083502 (2005).
\bibitem{Jedamzik06} K. Jedamzik, Phys. Rev. D 74, 103509 (2006).
\bibitem{Kusakabe09}M. Kusakabe, T. Kajino, T. Yoshida, T. Shima, Y. Nagai, and T. Kii, Phys. Rev. D 79, 123513 (2009).
\bibitem{Cyburt10} R. H. Cyburt, J. Ellis, B. D. Fields, F. Luo, K. A. Olive, and V. C. Spanos, J. Cosmol. Astropart. Phys. 10 032 (2010).
\bibitem{Voronchev12} V. T. Voronchev, Y. Nakao, K. Tsukida, and M. Nakamura, Phys. Rev. {\bf D85}, 067301 (2012).
\bibitem{Kusakabe14} M. Kusakabe, M.-K. Cheoun, and K. S. Kim, Phys. Rev. {\bf D90} 045009 (2014).
\bibitem{McDermott18} S. D. McDermott and M. S. Turner, arXiv:1811.04932.
\bibitem{Harigaya} K. Harigaya and K. Mukaida, J. High Energy Phys. (JHEP), {\bf 2014} 6 (2014).
\bibitem{Mukaida} K Mukaida and M Yamada, J. Cosmol Astropart. Phys (JCAP) {\bf 2016} 003 (2016).
\bibitem{Sasankan18} N. Sasankan, A. Kedia, M. Kusakabe and G. J. Mathews, arXiv:1810.05976 (2018).
\bibitem{Zenitani} S. Zenitani, Physics of Plasmas 22.4 (2015): 042116.
\bibitem{Melzani} M. Melzani, et al, Astronomy and Astrophysics 558 (2013): A133.
\bibitem{Meyer} J. P. Meyer, Astron. Astrophys. Suppl. Ser. 7, 417 (1972).
\bibitem{Boyd} R. N. Boyd, C. R. Brune, G. M. Fuller, and C .J. Smith, Phys. Rev. D 82, 105005 (2010).
\bibitem{Curado} E. MF. Curado, F. TL. Germani, and I. D. Soares, Physica A: Statistical Mechanics and its Applications 444 (2016): 963-969.
\bibitem{Landau-Lifshitz} L. D. Landau and E. M. Lifshitz, \textit{The Classical Theory of Fields}, Course of Theoretical Physics Vol. 2, 3rd ed. (Pergamon, Oxford, U.K., 1971).


\end{references}
\end{document}